\documentclass[prb,twocolumn,showpacs,amsmath,amssymb]{revtex4-1}
\usepackage{graphicx}
\newcommand{\beq}{\begin{equation}}
\newcommand{\eeq}{\end{equation}}
\newcommand{\beqa}{\begin{eqnarray}}
\newcommand{\eeqa}{\end{eqnarray}}

\begin{document}

\title{Possible mechanisms of electronic phase separation in oxide interfaces}


\author{N. Bovenzi$^1$,  F. Finocchiaro$^1$,  N. Scopigno$^1$,  D. Bucheli$^1$, S. Caprara$^1$, G. Seibold$^2$ and M. Grilli$^1$}

\affiliation{$^1$Dipartimento di Fisica, Universit\`a di Roma Sapienza, 
               piazzale Aldo Moro 5,  I-00185 Roma, Italy \\
$^2$Institut f\"ur Physik, BTU Cottbus-Senftenberg, 
 PBox 101344, 03013 Cottbus, Germany}

\begin{abstract}
 LaAlO$_3$/SrTiO$_3$ and  LaTiO$_3$/SrTiO$_3$ (LXO / STO) interfaces are known to host a strongly inhomogeneous (nearly) two-dimensional
 electron gas (2DEG). In this work we present three unconventional electronic mechanisms of electronic phase separation (EPS) in a 2DEG as a possible source of inhomogeneity in oxide interfaces. Common to all three mechanisms is the dependence of some (interaction) potential on the 2DEGÕs density. We first consider a mechanism resulting from a sizable density-dependent Rashba spin-orbit coupling. Next, we point that an EPS may also occur in the case of a density-dependent superconducting pairing interaction. Finally, we show that the confinement of the 2DEG to interface by a density-dependent, self-consistent electrostatic potential can by itself cause an EPS.
\keywords{Oxide interfaces \and Superconductivity \and Electronic phase separation}
\end{abstract}
\maketitle

\section{Introduction}
\label{intro}
The observation of a two-dimensional (2D) metallic state at the interface of two insulating oxides 
\cite{ohtomo,Mannhart:2008uj}, and the subsequent demonstration of its 
gate-tunable metal-to-superconductor transition 
\cite{reyren,triscone,espci1,espci2}, have attracted much 
attention in the last decade. Numerous experiments, like transport \cite{CGBC,BCCG,caprara,SPIN}, 
magnetometry \cite{ariando,luli,bert,metha1,bert2012}, tunneling \cite{richter,BCG}, 
and piezo-force spectroscopy \cite{fengbi},  indicate that the 2DEG is inhomogeneous. It seems likely that inhomogeneities 
at nanometric scales \cite{espciNM,fengbi} coexist with structural inhomogeneities at micrometric scales \cite{Kalisky}.
Although extrinsic sources like local or extended defects may contribute to this inhomogeneity, the dense nanoscopic
 character of the electronic charge clearly indicates that an intrinsic mechanism is at work leading to a nanoscopic EPS
  and to an inhomogeneous density distribution of the electrons.
In the present paper, we give an overview of three such mechanisms which could account for these generic features of the oxide 
interfaces. Based on the current state of knowledge it is hard to gauge the individual relevance of each mechanism (although they can work cooperatively). 
Therefore, we deem a systematic investigation worth being pursued.
The paper is structured as follows: In Sec. 2 we present a mechanism based on the density dependence of the Rashba spin-orbit coupling
(RSOC), which was found to be sizable in these systems. The second mechanism is based on the occurrence of superconductivity upon varying the 
electron density and is described in Sec. 3. In Sec. 4 we present the third mechanism based on the density dependence of the
electrostatic potential confining the electrons at the interface. Our final considerations are contained in Sec. 5.

\section{Electronic phase separation from Rashba spin-orbit coupling}
\label{sec:1}
We consider a 2d lattice model with strong RSOC described
by the hamiltonian
\begin{equation}                                         
H-t\sum_{\langle ij \rangle \sigma}c^\dagger_{i\sigma}c_{j\sigma} +H^{RSO} +\sum_{i,\sigma} \lambda_i \left\lbrack
c^\dagger_{i\sigma}c_{i\sigma} - n_i \right\rbrack . \label{eq:ham}
\end{equation}   
Here the first term describes the kinetic energy of electrons on a 
square lattice with nearest-neighbor ($\langle ij \rangle$) hopping.
The second term is the RSOC
\begin{equation}\label{eq:rso}
H^{RSO}=\sum_i \left\lbrack \gamma_{i,i+y} j_{i,i+y}^x - \gamma_{i,i+x} j_{i,i+x}^y
\right\rbrack
\end{equation}
where
$ j_{i,i+\eta}^\alpha = -i\sum_{\sigma\sigma'}\left\lbrack c^\dagger_{i\sigma} \tau^\alpha_{\sigma\sigma'} c_{i+\eta,\sigma'} - c^\dagger_{i+\eta,\sigma} \tau^\alpha_{\sigma'\sigma} c_{i,\sigma}
\right\rbrack
$
denotes the $\alpha$-component of the spin-current flowing on the bond
between $R_i$ and $R_{i+\eta}$.

\begin{figure}
\includegraphics[angle=0,scale=0.43]{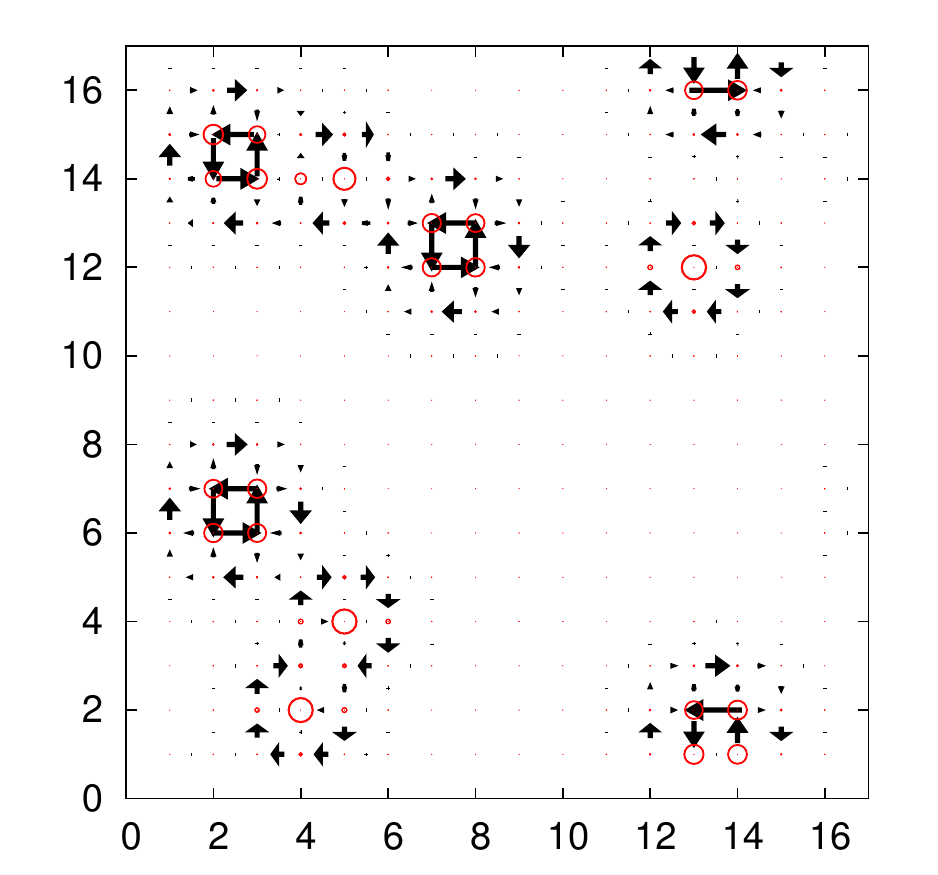}
\includegraphics[angle=0,scale=0.43]{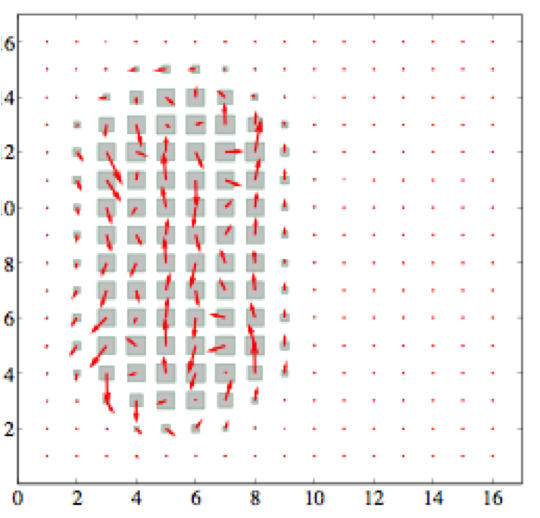}
%
\caption{Left panel: Charge distribution (red circles) and spin current
(z-component) in the ground state. Disorder potential $V_0=0.1t$, $n_{el}=26$ 
particles on a $16\times 16$ lattice with periodic boundaries. The RSOC is from Eq.(\ref{eq:coup})
with  $a_0=0.5$, $a_1=1.5$, $\beta_0 =\beta_1=0$. Right panel: Charge distribution (red circles) and current
(arrows) in the ground state for $N = 22$ electrons at $B = 1692 T$ and open boundary conditions with 
$a_0=0.5$, $a_1=1.5$, $\beta_0 =\beta_1=0$.
}
%
\label{SP-EPS}
\end{figure}

Following Ref. \cite{CPG} we assume that the coupling constants
depend on a perpendicular electric field $E$ which is proportional
to the local charge density. Since in real space the coupling constants
$\gamma_{i,i+\eta}$ are defined on the bonds, we discretize $E$ at the midpoints
of the bonds and define the dependence on the charge as
$
E_{i+\eta/2}=e_0 + e_1 (n_i + n_{i+\eta}).
$
For the dependence of the RSOC on the electric
field we adopt the form given in Ref. \cite{CPG}
so that altogether the following coupling is considered
\begin{equation}\label{eq:coup}
\gamma_{i,i+\eta}= \frac{a_0 + a_1 (n_i + n_{i+\eta})}{\lbrack 1+\beta_0 + \beta_1(n_i
+n_{i+\eta})\rbrack^3}. 
\end{equation}
For strong RSOC this coupling will induce the
formation of electronic inhomogeneities and thus concomitant
variations in the local chemical potential $\lambda_i$
obtained self-consistently by minimizing the energy with respect to the density:
\begin{equation}\label{eq:sc}
\lambda_i = \frac{\partial \gamma_{i,i+y}}{\partial n_i} \langle j_{i,i+y}^x\rangle
- \frac{\partial \gamma_{i,i+x}}{\partial n_i} \langle j_{i,i+x}^y\rangle .
\end{equation}
The left panel of Fig. \ref{SP-EPS} reports an example of EPS. A small amount of disorder is introduced to make the convergence to the inhomogeneous state easier, but in the absence of a density dependence of the RSOC it would have a minor effect leaving the system essentially homogeneous. In this latter case
the ground state is characterized by a homogeneous flow of $j^{x(y)}$ spin
currents along the y-(x)-direction whereas the z-component $j^z$ vanishes.  
On the other hand, the presence of a density dependent RSOC gives rise to a strongly inhomogeneous state which induces a finite flow of  $j^z$ spin currents 
(black arrows). These are caused by the in-plane electric fields 
$E_{\gamma=x,y}$ due to the inhomogeneous charge distribution 
which according to the response 
equation $j^{\alpha}_\beta=\sigma_s \varepsilon_{\alpha\beta\gamma}E_\gamma$ 
($\varepsilon_{\alpha\beta\gamma}$ denotes the totally antisymmetric tensor
and $\sigma_s$ the spin-Hall conductivity) 
generate $j^z$ spin currents along the $x-$ and $y-$directions. 
In general these currents are not conserved because the inhomogeneous 
RSOC introduces torque terms which act as source or drains of the 
spin current \cite{sonin}. Analogous non-conserved spin-currents bound to the
inhomogeneous charge density are observed for $j^{x}$ and $j^{y}$ (not shown).
The right panel of Fig. \ref{SP-EPS} displays an inhomogeneous state for the same model of Eq. (\ref{eq:ham}) in the presence of a strong perpendicular magnetic field.  The system is separated in a phase of density $n_2 \sim 0.31$ and the vacuum phase ($n_1 = 0$). 
Quite interestingly we found that in the quantum Hall regime the EPS instability is realized even for very small
RSOCs. The reason for this surprising result is that the EPS more easily occur in states with high density of states (DOS). This naturally favors the instability
for the Landau-Hofstadter states, which have a delta-like DOS. The fact that the (density-dependent) RSOC produces an inhomogeneous state
in the QH regime is an intriguing result that naturally calls for the investigation of edge states around and inside the inhomogeneous 
states\cite{nicandro}.

\section{Electronic phase separation from density dependent superconductivity}
\label{sec:2}
\subsection{Phenomenological Ginzburg-Landau model}
\begin{figure}
\includegraphics[angle=0,scale=0.26]{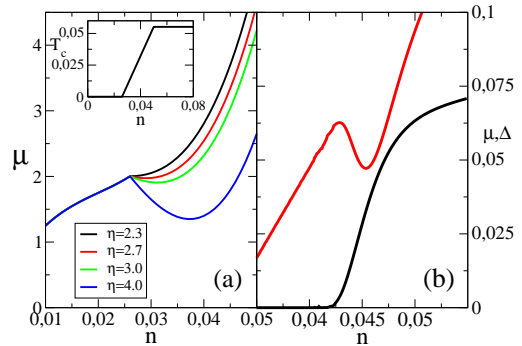}
%
\caption{(a) Chemical potential vs. density for various values of the parameter $\eta$  in $T_c(n)=\eta (n-n_c)$  and $n_c=0.026$ for the GL model;
(b) Chemical potential and SCOP as a function of the electron density $n$, with $n_c=0.04$, $a=0.1$ and $g_0=2.7t$ for the model of Eq.(\ref{pairing}).}
%
\label{GLmodel}
\end{figure}

A very interesting feature of LXO/STO interfaces is that superconductivity may be tuned by a gating potential which varies the density of the 2DEG.
Therefore these systems are characterized by a density dependent superconducting  critical temperature $T_c(n)$. 
We consider a standard Ginzburg-Landau (GL) model in which the superconducting order parameter (SCOP) is coupled to the density via the density dependence of the critical temperature. Assuming a spatially constant $\Delta$ one has
\begin{equation}
F\left[\Delta\right]=a\left[T-T_c(n)\right]
\Delta^2 + b\Delta^4.
\label{GLsc}
\end{equation}
Considering the simple form for $T_c(n)$ reported in the inset of 
Fig. \ref{GLmodel} (a), which starts linearly at
$n=n_c$, $T_c(n)=\eta (n-n_c)$, and then saturates to a constant value, one finds for the SCOP
$$
\Delta^2(n,T)=  \frac{a}{2b}(T_c(n)-T)\theta[T_c(n)-T] \,.
$$
In the absence of SC the chemical potential is a smoothly increasing function (positive electron compressibility) 
that can be obtained, e.g., from a generic van der Waals model for the 2DEG.
Taking the standard expression of the chemical potential $\mu=dF/dn$ one obtains that this smooth chemical potential
 is modified once $n\ge n_c$. Fig. \ref{GLmodel}(a) reports $\mu$ vs. $n$ for various values of the
parameter $\eta$. Clearly, if $\eta>\eta_c$ $\mu(n)$ acquires a negative slope, i.e. a negative compressibility, marking a PS
instability region, which must be found with a standard Maxwell construction.  Clearly the PS dome shrinks upon decreasing $\eta$
and vanishes at $n=n_c=0.026$

\subsection{Microscopic model with density dependent pairing}
The phenomenological Ginzburg-Landau approach can be complemented with 
a model, where electrons, described by a
tight-binding model (for simplicity we consider only a nearest-neighbor hopping $t$ on a square lattice), 
feel a pairing potential $g(n)$. A simple phenomenological form which reproduces a rapid growth of a SCOP above a
critical density can be introduced, e.g., as 
\begin{equation}
\label{pairing}
g(n)=g_0\tanh\left(\frac{n-n_c}{an} \right) \times\theta(n-n_c) \,.
\end{equation}
$g_0$ is a parameter controlling the maximum value attained by $g$ after its saturation and $a$ is a constant factor which determines how steep is the slope of $g(n)$ [and consequently of $\Delta(n)$]. At any fixed average density one can solve the coupled equations determining the SCOP (the standard BCS equation) and the chemical potential.
Fig. \ref{GLmodel} (b) reports the behavior of the SCOP and of the chemical potential $\mu$ as a function of the electron density near the critical density $n_c=0.04$ for $g_0=2.7t$. Interestingly, the chemical potential displays a non monotonic behavior, marking a region with negative compressibility starting in the proximity of the rapid growth
of the SCOP. Not only this finding agrees with the scenario emerging from the phenomenological  GL approach, but one finds that the EPS already occurs in a region of relatively weak superconducting coupling. Indeed, since the electronic tight-binding band has a DOS 
$N(E_F)\approx 1/8t$, the BCS coupling corresponding to $g_0=2.7t$ is in a regime of weak/moderate coupling $\lambda\sim 0.34$ at which the chemical potential barely starts to depend on $\lambda$. This finding is interesting because it indicates that the EPS mediated by a density-dependent pairing 
does not require strongly coupled electron pairs. 

\section{Electronic phase separation from interface confinement}
\label{sec:4}
\begin{figure}
\includegraphics[angle=0,scale=0.15]{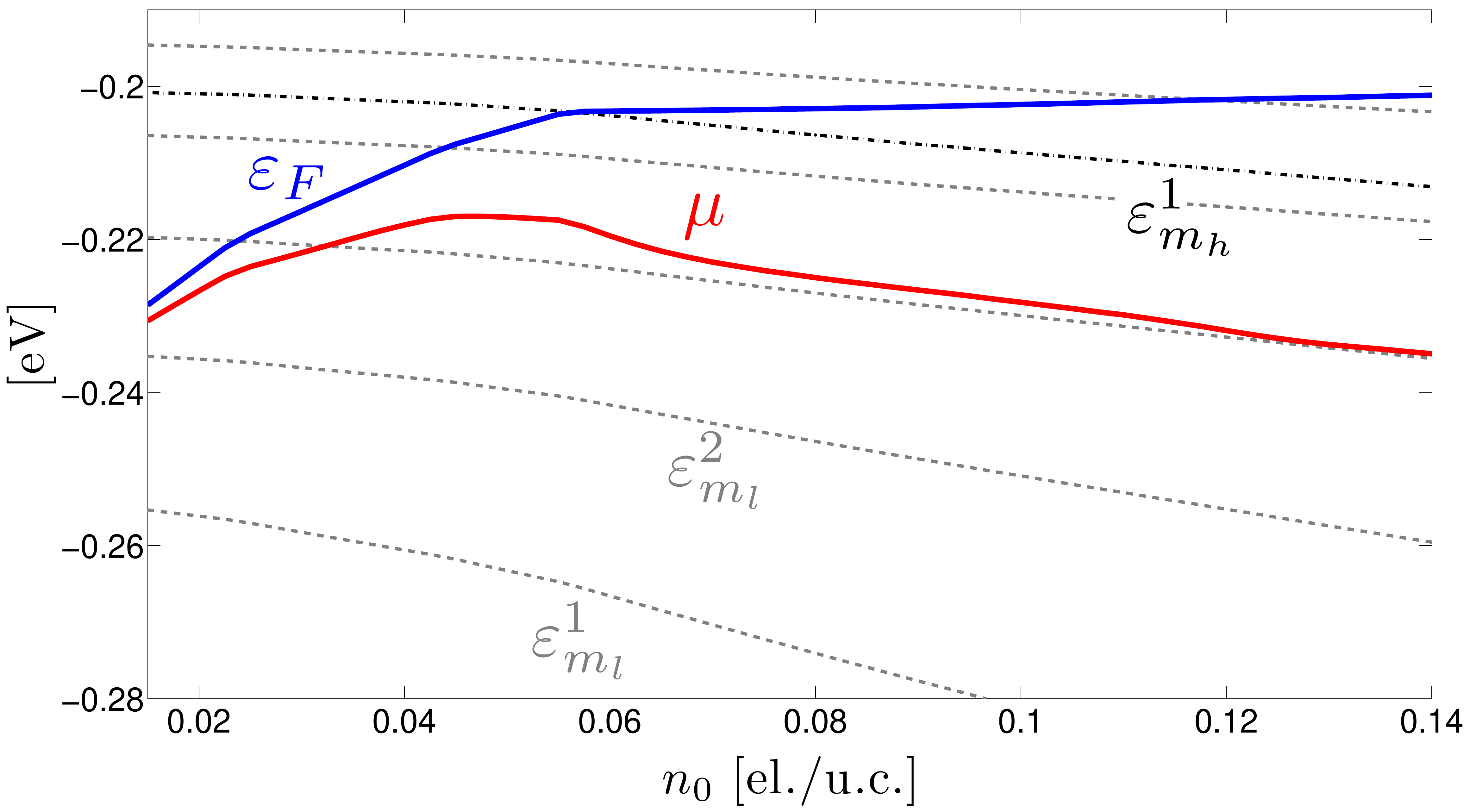}
\caption{Density dependence of the Fermi energy (solid blue curve), of the chemical potential (solid red curve), of the 
$d_{xy}$ levels (dashed grey curves), and of the $d_xz$ or $d_{yz}$ levels (dashed black curves).
}
%
\label{niccolo2}
\end{figure}

Another well known mechanism providing a negative contribution to the electronic compressibility is given by the mechanism confining the EG at interfaces. In this latter case, the self-consistent solution of the Schr\"odinger and Poisson equations relating the electronic wavefunctions and the electric potential arising from external potential and electronic density itself usually shows that the EG becomes more compressible once its finite transverse confinement is taken into account. The large dielectric constant of the material hosting the EG favors its ÒsofteningÓ, leading to an increased compressibility. In this regard, the STO, having a huge dielectric constant $\varepsilon >300$ is an optimal candidate to investigate
whether this mechanism may lead by itself to an overall negative compressibility and consequently to an EPS.
\begin{figure}
\includegraphics[angle=0,scale=0.3]{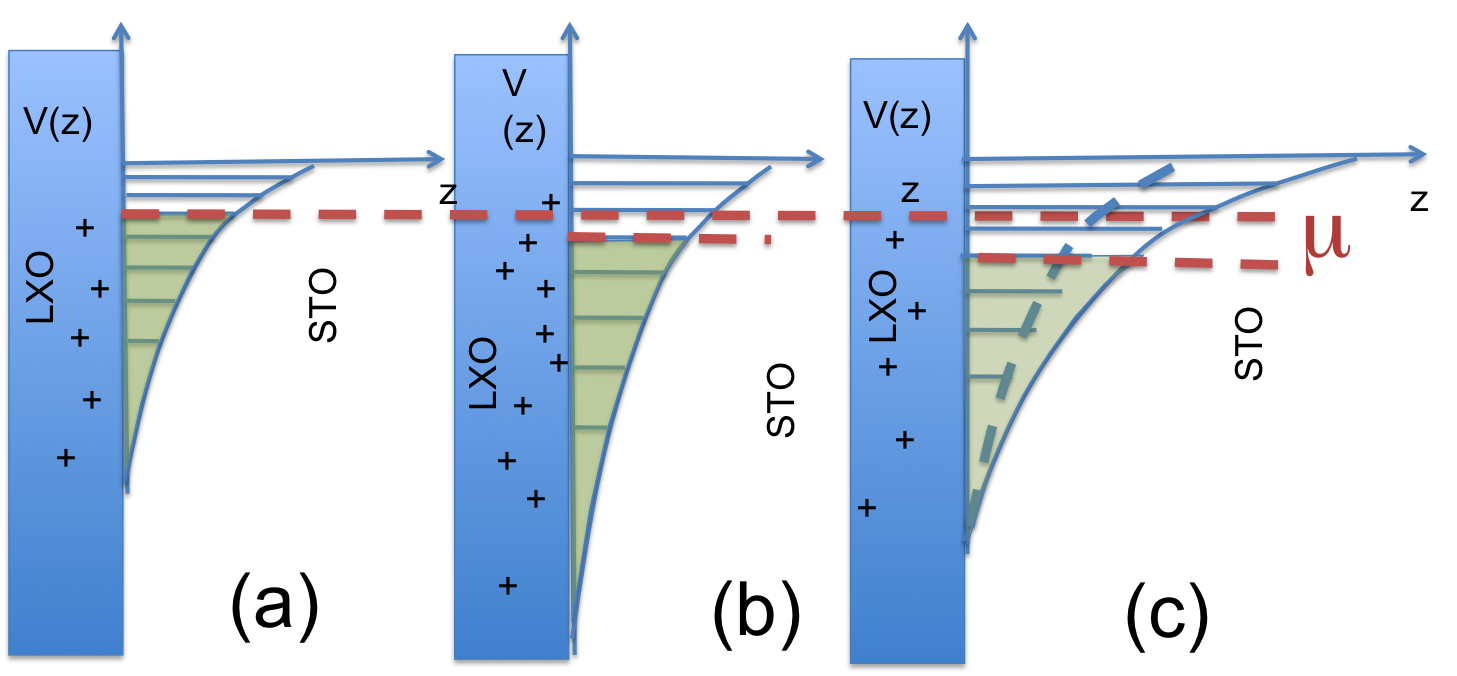}
%
\caption{(a) Schematic behavior of the selfconsistent electrostatic confining potential and of the sub-band electronic states perpendicularly to the LXO/STO interface. 
The LXO layer contains the positive countercharges (coming from the polarity catastrophe and/or from oxygen vacancies), while the occupied sub-band levels
are in marked in green. (b) When the density of electrons and countercharges is (locally) higher
the potential well becomes deeper (b) and when $d_{xz,yz}$ orbitals of Ti in STO start to be filled (c) the well also becomes broader. This induces
a decrease of the chemical potential $\mu$ (dashed red line).}
%
\label{niccolo1}
\end{figure}
Indeed, by numerically solving the coupled  Schr\"odinger and Poisson equations to determine the electronic density and sub-band structure,
we find that this is the case. Fig. \ref{niccolo1} (a) schematically displays the profile of the potential well along $z$ perpendicular to the LXO/STO interface.  Upon increasing the density, an increasing number of levels is filled, but the self-consistent electronic structure adjusts 
to lower its energy and, above some density, it displays a downward drift [Fig. \ref{niccolo1}(b)] and a broadening when sub-bands involving more $z$-oriented orbitals (like the $d_{xz}$ and $d_{yz}$ of Ti) [Fig.\ref{niccolo1}(c)]. These self-consistent modifications of the potential and of 
the electronic levels entail a pulling down of the chemical potential which eventually gives rise to a negative compressibility.
Fig. \ref{niccolo2} displays the electron density dependence of the sub-band Ti $t_{2g}$ levels, of the Fermi level,  and of the 
chemical potential. Notice that the compressibility stays negative [negative slope of $\mu(n)$] up to unphysically large densities. Of course several physical mechanisms will eventually stop this unphysical charge segregation (see below).
\section{Discussion and Conclusions}
\label{sec:5}
All the models outlined above are based on the common idea that some interactions lead to an energy decrease when the electron density is increased. In the presence of a RSOC $\gamma$ the bottom of the electronic band is lowered by a quantity $\epsilon_0\approx \gamma^2$. Thus, if $\gamma$ increases with $n$ as in Eq. (\ref{eq:coup}), there is an energetic benefit for the system to increase (locally) the density by segregating electrons. The same occurs in the model with density dependent pairing (\ref{pairing}): If the electronic density is increased above
the threshold $n_c$, electrons can be paired gaining binding (and possibly condensation) energy. In the electrostatic confinement model an energetic
gain occurs when the electrons and the positive countercharges in the LXO layer (those due to the polarity catastrophe and/or oxygen vacancies) are denser. 
In this case (cf. Fig. \ref{niccolo1} (b)) the electrons at the interface feel a deeper confining potential and lower their energy by phase separating.

All the above mechanisms of EPS are obviously contrasted by the Coulombic force that opposes the segregation of charged electrons.
However, two possibilities remain open to  allow for an inhomogeneous electronic distribution. On the one hand the electrons may separate on finite 
scales until the energetic gain is compensated by the cost of the Coulombic repulsion. Simple estimates \cite{noiPRB} show that the very large
value of the dielectric constant of STO weakens the Coulomb repulsion and allows this frustrated EPS mechanism to produce rather large 
($\sim 50$ nm) inhomogeneities. On the other hand, it is also possible that the positive countercharges (like the oxygen vacancies) diffuse
and follow the segregating electrons keeping charge neutrality. Of course also in this case EPS stops 
when the segregating electrons become too dense for the countercharges to follow, but finite inhomogeneities of substantial size can still 
be formed.





\end{document}